\documentstyle[12pt]{article}
\begin{document}
\parskip=6pt
\baselineskip=20pt
\bigskip
\centerline{\Large\bf Three-Dimensional Vertex Model in Statistical }
\centerline{\Large\bf Mechanics, from Baxter-Bazhanov Model\footnote
{This research was partially supported by the China center of advanced
science and technology.}}
\bigskip
\centerline{\large\bf Zhan-Ning HU \footnote{\bf email address:
huzn@itp.ac.cn}}
\smallskip
\centerline{ Institute of Theoretical Physics, Academia Sinica}
\centerline{ P. O. Box 2735, Beijing 100080, China\footnote{\bf mail address}}
\centerline{and}
\centerline{\large\bf Bo-Yu Hou}
\smallskip
\centerline{Institute of Modern Physics, Northwest University, Xian 710069,
China}
\vspace{2ex}
\bigskip
\begin{center}
\begin{minipage}{5in}
\centerline{\large\bf 	Abstract}
\vspace{1ex}
We find that the Boltzmann weight of the three-dimensional Baxter-Bazhanov
model is dependent on four spin variables which are the linear combinations
of the spins on the corner sites of the cube and the Wu-Kadanoff duality
between the cube and vertex type tetrahedron equations is obtained
explicitly for the Baxter-Bazhanov model. Then a three-dimensional vertex
 model is obtained by considering the symmetry property of the weight
function, which is corresponding to the three-dimensional Baxter-Bazhanov
 model. The vertex type weight function is parametrized as the dihedral
 angles between the rapidity planes connected with the cube. And  we write
down the symmetry relations of the weight functions under the actions of
the symmetry group $G$ of the cube. The six angles with a constrained
condition, appeared in the tetrahedron equation, can be regarded as the
 six spectrums connected with the six spaces in which the vertex type
tetrahedron equation is defined.
\bigskip

{\bf Keywords}:
Three-dimensional Baxter-Bazhanov model; Interaction-round-cube (IRC)
model; Wu-Kadanoff duality; Cube type tetrahedron equation; Vertex type
tetrahedron equation; Vertex type model; Three-dimensional lattice model;
 Symmetry property; Three-dimensional vertex model; Spherical trigonometry;
Vertex type weight function; Three-dimensional star-star relation;
 Additional constraints; Dihedral angles; symmetry group.
\end{minipage}
\end{center}
\newpage

\baselineskip=30pt

\section{\bf Introduction}

Recently big progress has been made in three-dimensional integrable models
 in statistical mechanics. Bazhanov and Baxter \cite{BB1} introduced the
 Interaction-Round-a-Cube (IRC) model which is the generalization of $N=2$
Zamolodchikov model \cite{Zam1}. Kashaev $et~al$ \cite{Kaev2} showed that
 the Boltzmann weights of the Baxter-Bazhanov model satisfy the cube type
 tetrahedron equation by introducing  the star-square relation for which
 the connection is found \cite{humod} with the choral Potts model. The
 restricted star-triangle relation and the star-star relation  of this
 model have been discussed in detail in Refs \cite {BB2,Kaev1,hu1,huJsta}.
And they connected with the quantum dilogarithm \cite{kaenew} and the
 shift operator in discrete space-time picture \cite{huphy,hueur}. Then
 the new series of the three-dimensional integrable lattice models were
 presented by Mangazeev $et~al$ \cite{Newseies} of which the weight
functions satisfy modified tetrahedron equation \cite{elliptic}. Recently,
 Cerchiai $et~al$ studied the Baxter-Bazhanov model from the point of link
 theory and given the representations of the braid group if some suitable
 spectral limits are taken.

In red. \cite{Kor} Korepanov got the solution of vertex tetrahedron
equation with the spin variables taking $N=2$ values, which leads to a
commuting family of transfer-matrices. From the respect of the scattering
 process Hietarinta discussed the three corresponding tetrahedron equations
 in which the Frenkel-Moore equation was fitted \cite{FreMor,leehu2} and
 proposed another vertex solution with 16 nonzero weights \cite{Hira}.
 And the discrete symmetry groups of vertex models were studied by
Boukraa $et~al$. As a generalization of Hietarinta's solution of
tetrahedron equation  Mangazeev $et~al$ \cite{newsolu} proposed another
 $N$-state spin integrable model on a three-dimensional lattice and this
 model can be reformulated as a vertex model. Now we know that the weight
 function of this model can be obtained from Baxter-Bazhanov model by
taking some limits \cite{hunew}. It is naturally  to ask if a
three-dimensional vertex model exists corresponding to three-dimensional
 Baxter-Bazhanov model. One of the aims of this paper is to give a
positive answer. And we showed that  the weight function of the
 Baxter-Bazhanov model is dependent on four spin variables which are the
 linear combinations of the spins located on the corner sites of the
 elementary cube. The duality between cube and vertex type weight
functions is obtained explicitly. The spectrums have the symmetrical
 property in respect of vertex forms.

This paper is organized as the follows. In section 2 we give a brief
describe of the Baxter-Bazhanov model and the duality between the cube
 and the vertex type tetrahedron equations. The weight functions of the
Baxter-Bazhanov model are written as the vertex forms in section 3.
Then the duality is obtained explicitly for the Baxter-Bazhanov model.
 By using the symmetry properties of the weight functions we get the
 vertex type weight functions for the three-dimensional vertex model.
 It should be noted that the weight function of model proposed by
Mangazeev $et~al$ can be obtained from this vertex type weight function
when we take the limit of the spectrum and  use the star-triangle relation
 of the Baxter-Bazhanov model. In section 4 the vertex type weight
 functions are parametrized as the angles of the spherical triangles by
using the methods of the spherical trigonometry parametrization. These
angles are the dihedral angles between the ``rapidity planes'' passing
 the cubes similarly as in the Zamolodchikov model. In this way, the
 spectrums appeared in the vertex type tetrahedron equation can be
denoted by these angles and they connect with the spaces in which the
vertex type tetrahedron equation is defined. In section 5 we discuss the
 constrained conditions imposed on the tetrahedron equations from the point
of the angle variables. Then the symmetry properties of the vertex type
weight functions are discussed. They are symmetrical about the
transformations of the group $G$ consisting of various rotations,
 reflections and their combinations of the cube. Finally, some
conclusions and remarks are given.

\section{\bf Baxter-Bazhanov Model and Duality between Cube and Vertex
Type Tetrahedron Equation }

\subsection{ Three-Dimensional Baxter-Bazhanov Model }

As is well known, the Baxter-Bazhanov model is an
Interaction-Round-a-Cube (IRC) Model. The partition function of it reads
\begin{equation}
Z=\sum_{spins}\prod_{cubes}W(a|efg|bcd|h)
\end{equation}
where $W(a|efg|bcd|h)$ is the Boltzmann weight of the spin configuration
$a,\cdots,h$ (see Fig. 1.) and these spin variables take their values in
 $Z_N$ with $N\ge2$. The product is over all elementary cubes in the
 simple cubic lattice ${\cal L}$. The Boltzmann weight $W(a|efg|bcd|h)$
 can be written as
$$
W(a|efg|bcd|h)
{}~~~~~~~~~~~~~~~~~~~~~~~~~~~~~~~~~~~~~~~~~~~~~~~~~~~~~~~~~~~~~~~~~~~~
{}~~~~~~~~~~~~
$$$$
={w(v_4/v_3,e-c-d+h)s(g,a-g-f+b)\over w(v_4/v_3,a-g-f+b)s(c-h,h-d)} ~~~~~~
{}~~~~~~~~~~
$$
\begin{equation} \label{1}
{}~~~~~~~\times \Bigg\{ \sum^{N-1}_{\sigma=0}{w(v_2,b-f+\sigma)
w(v_3,d-h-\sigma)s(\sigma,a)s(\sigma,h)\over w(v_1,g-a+\sigma)
w(v_4,e-c-\sigma)s(\sigma,c)s(\sigma,f)}\Bigg\}_0,
\end{equation}
with the relation $\omega v_1v_4=v_2v_3$, where  the subscript "0" after
 the curly brackets indicates that the  expression in the braces is divided
by itself with the zero exterior spins and  we have used the notations
\begin{equation}
{w(v,a)\over w(v,0)}=[\Delta(v)]^a\prod^a_{j=1}(1-\omega^jv)^{-1},
{}~~\Delta(v)=(1-v^N)^{1/N}
\end{equation}
\begin{equation}
\omega=\exp(2\pi i/N),~~ \omega^{1/2}=\exp(\pi i/N),~~s(a,b)=\omega^{ab}
\end{equation}
Note that the Boltzmann weight function (\ref{1}) describes a very special
 type of the interaction of eight spins around the cube as in Fig.1.
 There are three-spin interactions on the triangles $(a,g,\sigma),
 (b,f,\sigma), (d,h,\sigma), (c,e,\sigma)$, described by $w(v,a)$ or by
 $1/w(v,a)$, and two-spin interactions $s(\sigma,a), s(\sigma,c),
 s(\sigma,h), s(\sigma,f)$ associated with the edges linking $\sigma$
to $a,c,f,h$ in the curly brackets . The factors before the curly
brackets denote the spins interactions in the planes $(a,f,b,g)$ and
 $(c,e,d,h)$. After introducing an overall normalization and some
 additional multipliers \cite{Kaev1}, the weight function of the
 Baxter-Bazhanov model is expressed as
$$
W_P(a|efg|bcd|h) ~~~~~~~~~~~~~~~~~~~~~~~~~~~~~~~~~~~~~~~~~~~~~~~~~~~~
{}~~~~~~~~~~~~
$$$$
=\frac{\omega^{fb}}{\omega^{ag}}\Bigg[\frac{w(x_{14}x_{23},x_{12}x_{34},
x_{13}x_{24}|a+d,e+f)}{w(x_{14}x_{23},x_{12}x_{34},x_{13}x_{24}|g+h,c+b)}
\Bigg]^{1/2}~~~
$$$$
\times \Bigg[\frac{w(x_4,x_{34},x_3|e+h,d+c)}{w(x_4,x_{34},x_3|a+b,f+g)}
\Bigg]^{1/2} ~~~~~~~~~~~~~~~~~~
$$$$
\times \Bigg[\frac{w(x_2,x_{12},x_1|e+g,a+c)}{w(x_2,x_{12},x_1|d+b,f+h)}
\Bigg]^{1/2}\frac{\omega^{(ag+gb+bh)/2}}{\omega^{(hd+de+ea)/2}}~
$$
\begin{equation} \label{W1}
{}~~~~~~~~~~~~~\times \Bigg\{\sum_{\sigma \in Z_N}
\frac{w(x_3,x_{13},x_1|d,h+\sigma)w(x_4,x_{24},x_2|a,g+\sigma)
}{w(x_4,x_{14},x_1|e,c+\sigma)w(x_3/\omega,x_{23},x_2|f,b+\sigma)}
\Bigg\}_0
\end{equation}
It satisfies the tetrahedron equation which ensures the commutativity
 of the layer-to-layer transfer matrices. Here we have used the function
\begin{equation}
w(x,y,z|k,l)=w(x,y,z|k-l)\Phi(l),~~w(x,y,z|l)=\prod^{l}_{j=1}\frac{y}
{z-x\omega^j},~~k,l\in Z_N
\end{equation}
with the notation
\begin{equation}
x^N+y^N=z^N,~~\Phi(l)=\omega^{l(l+N)/2},~~x^N_i-x^N_j=x^N_{ij},
\end{equation}
for $i<j$ and $i,j=1,2,3,4$.

\subsection{ Duality between Cube and Vertex Type Tetrahedron Equation }

The Boltzmann weight function $W$ in Eq.(\ref{W1}) satisfies the
following tetrahedron equation \cite{Kaev2}
$$
\sum_{d}
W(a_4|c_2c_1c_3|b_1b_3b_2|d)W'(c_1|b_2a_3b_1|c_4dc_6|b_4)  ~~~~~~~~~~~~
{}~~~~~~~~~~~~~~~~~~~~
$$$$
\times W''(b_1|dc_4c_3|a_2b_3b_4|c_5)W'''(d|b_2b_4b_3|c_5c_2c_6|a_1)~~~~~
{}~~~~~~~~~~~~~~~~~~
$$$$
{}~~~~~~~~~~~~~~~~~=\sum_{d}
W'''(b_1|c_1c_4c_3|a_2a_4a_3|d)W''(c_1|b_2a_3a_4|dc_2c_6|a_1)
$$
\begin{equation} \label{cube}
{}~~~~~~~~~~~~~~~~~~~~~~~~~~~\times W'(a_4|c_2dc_3|a_2b_3a_1|c_5
)W(d|a_1a_3a_2|c_4c_5c_6|b_4)
\end{equation}
where $W$, $W'$, $W''$ and $W'''$ are some four
 sets of Boltzmann weights. From the respect of the scattering process
and using the particle labelling schemes, J. Hietarinta \cite{Hira}
 written down the vertex type tetrahedron equation \cite{Kor,newsolu}
\begin{equation} \label{vertex}
{\displaystyle\sum_{k_1,k_2,k_3,\atop k_4,k_5,k_6}}
R^{k_1,k_2,k_3}_{i_1,i_2,i_3}R'^{j_1k_4k_5}_{\phantom{,}k_1i_4i_5}
R''^{j_2j_4k_6}_{\phantom{,,}k_2k_4i_6}
R'''^{j_3j_5j_6}_{\phantom{,,,}k_3k_5k_6}
={\displaystyle\sum_{k_1,k_2,k_3,\atop k_4,k_5,k_6}}
R'''^{k_3,k_5,k_6}_{\phantom{,,,}i_3,i_5,i_6}
R''^{k_2k_4j_6}_{\phantom{,,}i_2i_4k_6}
R'^{k_1j_4j_5}_{\phantom{,}i_1k_4k_5}
R^{j_1j_2j_3}_{k_1k_2k_3}
\end{equation}
Then we call the relation (\ref{cube}) as the cube type tetrahedron
 equation. Just as the Wu-Kadanoff duality in the Yang-Baxter equation,
 the tetrahedron analogue of the Wu-Kadanoff duality between the above
 two type tetrahedron equations can be constructed by
\begin{equation}
W(a|efg|bcd|h)=R^{\alpha d+\beta h+\gamma b+\delta f,\alpha
 h+\beta c+\gamma g+\delta b,\alpha d+\beta e+\gamma c+
\delta h}_{\alpha e+\beta c+\gamma g+\delta a,\alpha d+\beta e+
\gamma a+\delta f,\alpha f+\beta a+\gamma g+\delta b}
\end{equation}
where the constants $\alpha, \beta, \gamma, \delta$ are the parameters
 of the map $F_{cv}: W=R\circ F_{cv}$ (see Ref.\cite{Hira}). There are
 two non-trivial results about the map $F_{cv}$:
\begin{equation} \label{2.2}
R^{lmn}_{ijk}=0 ~~\mbox{unless} ~~\alpha l+\beta i=\beta m+\alpha j
 ~~\mbox{and}~~\gamma m+\beta j=\beta n+\gamma k
\end{equation}
for the case of $\alpha\gamma=\beta\delta$, and
\begin{equation} \label{2.3}
R^{lmn}_{ijk}=0 ~~\mbox{unless} ~~m=i+k ~~\mbox{and}~~j=l+n
\end{equation}
for the case of $\alpha=\gamma=0$ and $\beta=-\delta=1$. The solution
presented in Ref. \cite{newsolu} corresponds to the latter, which can
be obtained from the Boltzmann weight of the Baxter-Bazhanov model by
taking some limits\cite{hunew}. In the following section the map $F_{cv}$
will be obtained for the three-dimensional Baxter-Bazhanov model. Then we
 get a three-dimensional vertex model which corresponds to the IRC model.

\section{\bf Three-Dimensional Vertex Model}

\subsection{The Vertex Type Boltzmann Weight}
{}From the expressions (\ref{1}) and (\ref{W1}), we know that the Boltzmann
weights of the Baxter-Bazhanov model are dependent on the eight spins
located on the corner sites of the elementary cube. That is, the Boltzmann
weights map as $W:~ I^{\otimes 8}_c\rightarrow \cal{C}$ and the cube
 type tetrahedron equation is defined on $I^{\otimes 15}_c$. The
 relations (\ref{2.2}) and (\ref{2.3}) mean that two labels of $R$ are
 determined from the others. What is happened about the
 three-dimensional Baxter-Bazhanov model?  Now we deal with it. Set
\begin{equation}
r_1=d-h,~ r_2=a-g,~ r_3=e-c,~r_4=f-b,~r_5=g+h-b-c
\end{equation}
Using the relation (6) we can change the expression (5) into the form:
$$
W_P(a|efg|bcd|h) ~~~~~~~~~~~~~~~~~~~~~~~~~~~~~~~~~~~~~~~~~~~~~~~~~~~~~
{}~~~~~~~~~~~
$$$$
=I_\omega\Bigg[\frac{w(x_{14}x_{23},x_{12}x_{34},x_{13}x_{24}|r_1+
r_2-r_3-r_4+r_5)}{w(x_{14}x_{23},x_{12}x_{34},x_{13}x_{24}|r_5)}~~~~~
$$$$
\times \frac{w(x_4,x_{34},x_3|r_3-r_1)w(x_2,x_{12},x_1|r_3-r_2)}
{w(x_4,x_{34},x_3|r_2-r_4)w(x_2,x_{12},x_1|r_1-r_4)}\Bigg]^{1/2} ~~~
$$
\begin{equation} \label{W2}
{}~~~~~~~~~~~~~~~~~\times \Bigg\{\sum_{\sigma \in Z_N}\frac{w(x_3,x_{13},
x_1|r_1+\sigma)w(x_4,x_{24},x_2|r_2+\sigma)}{\omega^{\sigma r_5}
w(x_4,x_{14},x_1|r_3+\sigma)w(x_3/\omega,x_{23},x_2|r_4+\sigma)}\Bigg\}_0
\end{equation}
where
\begin{equation}
I_{\omega}=(-)^{r_5}(\omega^{1/2})^{r^{~2}_3+r_3r_4-r_1r_3-r_2r_3-r_3
r_5-r_4r_5}\Bigg[\frac{\Phi(r_1)\Phi(r_2)}{\Phi(r_3)\Phi(r_4)}\Bigg]^{1/2}
\end{equation}
$\Phi(r_i),~i=1,2,3,4,$ are given in relation (7). By taking account
 of the property
\begin{equation}
w(x,y,z|l)w(z,\omega^{1/2}y,\omega x|-l)\Phi(l)=1,\quad l\in Z_N
\end{equation}
the weight function becomes
 $$
W_P(a|efg|bcd|h)=(-)^{k_2}(\omega^{1/2})^{k_1k_2+k_2k_3+k_1k_3}
\Bigg[\frac{w(x_1,\omega^{1/2}x_{12},\omega x_2|k_1)}
{w(x_1,\omega^{1/2}x_{12},\omega x_2|k_4-k_3)}    ~~~~~~
$$$$
{}~~~~~~~~~~~~\times\frac{w(x_{14}x_{23},x_{12}x_{34},x_{13}x_{24}
|k_4-k_1-k_2-k_3)w(x_4,x_{34},x_3|k_3)}
{w(x_{14}x_{23},x_{12}x_{34},x_{13}x_{24}|k_2)
w(x_4,x_{34},x_3|k_4-k_1)}\Bigg]^{1/2}
$$
\begin{equation} \label{W3}
\times \Bigg\{\sum_{\sigma \in Z_N}\frac{\omega^{\sigma k_2}
w(x_3,x_{13},x_1|\sigma)w(x_4,x_{24},x_2|k_4+\sigma)}
{w(x_4,x_{14},x_1|k_3+\sigma)w(x_3/\omega,x_{23},x_2|k_1+\sigma)}\Bigg\}_0
\end{equation}
where
\begin{equation}
k_1=r_4-r_1,~k_2=-r_5,~k_3=r_3-r_1,~k_4=r_2-r_1
\end{equation}
So the Boltzmann weight of the Baxter-Bazhanov model can be reformulated as
$$
R^{j_1j_2j_3}_{i_1i_2i_3}=(-)^{j_2}(\omega^{1/2})^{j_1j_2+j_2j_3+j_1j_3}
{}~~~~~~~~~~~~~~~~~~~~~~~~~~~~~~~~~~~~~~~~~~~~~~~~~~~~~~~~~~~
$$$$
{}~~~~~~~~~~\times\Bigg[\frac{w(x_1,\omega^{1/2}x_{12},\omega x_2|j_1)
w(x_{14}x_{23},x_{12}x_{34},x_{13}x_{24}|-i_2)w(x_4,x_{34},x_3|j_3)}
{w(x_1,\omega^{1/2}x_{12},\omega x_2|i_1)
w(x_{14}x_{23},x_{12}x_{34},x_{13}x_{24}|-j_2)w(x_4,x_{34},x_3|i_3)}
\Bigg]^{1/2}
$$
\begin{equation} \label{R1}
\times \Bigg\{\sum_{\sigma \in Z_N}\frac{\omega^{\sigma j_2}
w(x_3,x_{13},x_1|\sigma)w(x_4,x_{24},x_2|i_1+j_3+\sigma)}
{w(x_4,x_{14},x_1|j_3+\sigma)w(x_3/\omega,x_{23},x_2|j_1+\sigma)}
\Bigg\}_0  ~~~~~~~~~~~~
\end{equation}
where the spin variables $i_1$, $i_2$, $i_3$, $j_1$, $j_2$, $j_3$
 satisfy
 the conditions $i_1+i_2=j_1+j_2,~ i_2+i_3=j_2+j_3$ and
$$
i_1=a+c-e-g,~~i_2=e+f-a-d,~~i_3=a+b-f-g
$$
\begin{equation}
j_1=f+h-b-d,~~j_2=b+c-g-h,~~j_3=e+h-c-d
\end{equation}
Comparing with the relation (10) we know that the parameters of the
 map $F_{cv}$ are $\alpha=-\beta=\gamma=-\delta=-1$. The expression
 (\ref{R1}) can be interpreted as the Boltzmann weight of a
three-dimensional vertex model. The Boltzmann weight of the vertex
model proposed in Ref. \cite{newsolu} can be obtained when we set
 $i_3=j_3=0$ and use the star-triangle relation of the Baxter-Bazhanov
 model (see Ref. \cite{hunew}).

 \subsection{The Spectral Parameters in Weight Function}
 From the symmetry properties of the Boltzmann weights of the
 Baxter-Bazhanov model , we have the relation
$$
\Bigg\{\sum_{\sigma\in Z_N}\frac{w(x_3,x_{13},x_1|\sigma +a)
w()x_4,x_{24},x_2|\sigma+c)s(\sigma,n)}{w(x_4,x_{14},x_1|\sigma
 +b)w()x_3/\omega,x_{23},x_2|\sigma+d)}\Bigg\}_0 ~~~~~~~~~~~~~~~
{}~~~~~~~~~~~~~~~~~~
$$$$
=\frac{w(x_4,x_{34},x_3|c-d)}{w(x_4,x_{34},x_3|b-a)s(a,n)} ~~~~~~~
{}~~~~~~~~~~~~~~~~~~~~~~~~~~~~~~~~~~~~~~~~~~~~
$$
\begin{equation}
{}~~~~~~~~~~~~~\times \Bigg\{\sum_{\sigma\in Z_N}
\frac{w(x_4x_{13},x_1x_{34},x_3x_{14}|\sigma-a+b+n)
w(x_{23},x_{34},x_{24}|\sigma)s(\sigma,d)}
{w(x_{13},\omega x_{34},\omega x_{14}|\sigma+n)
w(x_4x_{23},x_2x_{34},x_3x_{24}|\sigma+c-d)s(\sigma,a)}\Bigg\}_0
\end{equation}
where $a,b,c,d,\sigma,n \in Z_N$ (see Refs. \cite{Kaev2,hu1}). By using
the above relation the vertex type weight function (\ref{R1}) can be
written as
$$
R^{j_1j_2j_3}_{i_1i_2i_3}=(-)^{j_2}(\omega^{1/2})^{j_1j_2+j_2j_3+j_1j_3}
 ~~~~~~~~~~~~~~~~~~~~~~~~~~~~~~~~~~~~~~~~~~~~~~~~~~~~~~~~~~~
$$$$
{}~~~~~~~~~~\times\Bigg[\frac{w(x_1,\omega^{1/2}x_{12},\omega x_2|j_1)
w(x_{14}x_{23},x_{12}x_{34},x_{13}x_{24}|-i_2)w(x_4,x_{34},x_3|i_3)}
{w(x_1,\omega^{1/2}x_{12},\omega x_2|i_1)
w(x_{14}x_{23},x_{12}x_{34},x_{13}x_{24}|-j_2)w(x_4,x_{34},x_3|j_3)}
\Bigg]^{1/2}
$$
\begin{equation} \label{R2}
{}~~~~~~~~~~~~~\times \Bigg\{\sum_{\sigma\in Z_N}
\frac{w(x_4x_{13},x_1x_{34},x_3x_{14}|\sigma+j_2+j_3)
w(x_{23},x_{34},x_{24}|\sigma)s(\sigma,j_1)}{w(x_{13},\omega x_{34},
\omega x_{14}|\sigma+j_2)w(x_4x_{23},x_2x_{34},x_3x_{24}|\sigma+i_3)}
\Bigg\}_0 ~~
\end{equation}
Set
\begin{equation}
u=\frac{x_1}{\omega x_2},~ v=\frac{x_4}{x_3},~z=\frac{z_1}{z_2},~z_1=
\frac{x_{13}}{\omega x_{14}},~z_2=\frac{x_{23}}{x_{24}}
\end{equation}
The Boltzmann weight of the three-dimensional vertex model showed in
Fig.2 has the form:
$$
R(u,z,v)^{j_1j_2j_3}_{i_1i_2i_3}=(-)^{j_2}
(\omega^{1/2})^{j_1j_2+j_2j_3+j_1j_3}\Bigg[
\frac{w(u,j_1)w(z_2/(\omega z_1),-i_2)w(v,i_3)}{w(u,i_1)
w(z_2/(\omega z_1),-j_2)w(v,j_3)}\Bigg]^{1/2}
$$
\begin{equation}
\times\Bigg\{\sum_{\sigma\in Z_N}\frac{w(\omega vz_1,\sigma+j_2+j_3)
w(z_2,\sigma)s(\sigma,j_1)}{w(z_1,\sigma+j_2)w(vz_2,\sigma+i_3)}\Bigg\}_0
\end{equation}
where we have used the notation (3). It satisfies the vertex type
 tetrahedron equation
$$
{\displaystyle\sum_{\{k_i\},\atop i=1,\cdots,6}}
R(u_1,u_2,u_3)^{k_1,k_2,k_3}_{i_1,i_2,i_3}
R(u_1,u_4,u_5)^{j_1k_4k_5}_{k_1i_4i_5}
R(u_2,u_4,u_6)^{j_2j_4k_6}_{k_2k_4i_6}
R(u_3,u_5,u_6)^{j_3j_5j_6}_{k_3k_5k_6}= ~~~~~~~
$$
\begin{equation}
{}~{\displaystyle\sum_{\{k_i\},\atop i=1,\cdots,6}}
R(u_3,u_5,u_6)^{k_3,k_5,k_6}_{i_3,i_5,i_6}
R(u_2,u_4,u_6)^{k_2k_4j_6}_{i_2i_4k_6}
R(u_1,u_4,u_5)^{k_1j_4j_5}_{i_1k_4k_5}
R(u_1,u_2,u_3)^{j_1j_2j_3}_{k_1k_2k_3}
\end{equation}
where
$$
u_1=\frac{x_1}{\omega x_2}=\frac{x_1'}{\omega x_2'},~~~~~~~~~~~~u_2
=\frac{x_{13}x_{24}}{\omega x_{14}x_{23}}=\frac{x_1''}{\omega x_2''},
{}~~u_3=\frac{x_4}{x_3}=\frac{x_1'''}{\omega x_2'''}
$$
\begin{equation}
u_4=\frac{x_{13}'x_{24}'}{\omega x_{14}'x_{23}'}=\frac{x_{13}''x_{24}''}
{\omega x_{14}''x_{23}''},~~u_5=\frac{x_4'}{x_3'}=\frac{x_{13}'''x_{24}'''}
{\omega x_{14}'''x_{23}'''},~~~~u_6=\frac{x_4''}{x_3''}=\frac{x_4'''}
{x_3'''}~
\end{equation}
The other constraints on the spectrums will be discussed in the
 section 5. We can think of  each side of the cube type tetrahedron
 equation as the partition function of the four skewed cubes joined
together with  a common interior spin $d$, which forms a rhombic
dodecahedron. In this way, we can express both of the two type tetrahedron
 equations in Figs. 3, 4, graphically. These figures give also the duality
 between the cube type and vertex type tetrahedron equations.

\section{\bf The Spectral Parametrization by Using the Spherical Trigonometry}

In this section we parametrize the spectrums of the Boltzmann weights as
the dihedral angles between the ``rapidity planes'' passing the cubes
similarly as in the Zamolodchikov model \cite{Zam1}. Following the methods
 in Refs. \cite{Kaev1,BB2}, we introduce a large sphere (its radius is
 much larger than the size of the tetrahedra) with a point near the
 vertices as the center. Consider four great circles on the sphere
corresponding to the four ``world planes'' (see Ref. \cite{Zam1}). A
 fragment of the stereo-graphic projection of this sphere is shown in
 Fig. 5. Note that we use  the angles which is different from  the ones
 of Zamolodchikov's. Define
\begin{equation}\begin{array}{lll}
l_1=l_{23}/N,&l_2=l_{13}/N,&l_3=l_{12}/N\\[1.5mm]
l_1'=l_{45}/N,&l_2'=l_{15}/N,&l_3'=l_{14}/N\\[1.5mm]
l_1''=l_{46}/N,&l_2''=l_{26}/N,&l_3''=l_{24}/N\\[1.5mm]
l_1'''=l_{56}/N,&l_2'''=l_{36}/N,&l_3'''=l_{35}/N\\[1.5mm]
\end{array}\end{equation}
where $l_{ij} (i,j=1,\cdots,6, i<j)$ denotes the length of the segment
between $i$ and $j$ along the circle. Then we can write down
\begin{equation}\begin{array}{ll}
x_1=c_1/s_1,&x_2=\omega^{-1/2}s_1/c_1\\[1.5mm]
x_3=exp(-il_2)s_3/c_3,&x_4=\omega^{-1/2}exp(-il_2)c_3/s_3\\[1.5mm]
x_{12}=1/(c_1s_1),&x_{13}=exp[i(l-l_2)]c_2/(c_3s_1)\\[1.5mm]
x_{14}=exp[i(l_3-l)]s_2/(s_1s_3),&x_{23}=\omega^{-1/2}exp[i(l_1-l)]
s_2/(c_1c_3)\\[1.5mm]
x_{24}=exp(-il)c_2/(s_3c_1),&x_{34}=exp(-il_2)/(c_3s_3)
\end{array}\end{equation}
And the primes added to the $x$'s are  correspondent to that of the
$c_i,s_i,l_i,l$ with
\begin{equation}\begin{array}{ll}
l=(l_{12}+l_{13}+l_{23})/(2N),&l'=(l_{14}+l_{15}+l_{45})/(2N)\\[1.5mm]
l''=(l_{24}+l_{26}+l_{46})/(2N),&l'''=(l_{35}+l_{36}+l_{56})/(2N)\\
\end{array}\end{equation}
\begin{equation}\begin{array}{ll}
c_1'=c_1=[cos(\frac{\displaystyle{\theta_1}}{\displaystyle{2}})]^{1/N},
&s_1'=s_1=[sin(\frac{\displaystyle{\theta_1}}{\displaystyle{2}})]^{1/N}\\[2mm]
c_1''=c_2=[cos(\frac{\displaystyle{\theta_2}}{\displaystyle{2}})]^{1/N},
&s_1''=s_2=[sin(\frac{\displaystyle{\theta_2}}{\displaystyle{2}})]^{1/N}\\[2mm]
c_1'''=c_3=[cos(\frac{\displaystyle{\theta_3}}{\displaystyle{2}})]^{1/N},
&s_1'''=s_3=[sin(\frac{\displaystyle{\theta_3}}{\displaystyle{2}})]^{1/N}\\[2mm]
c_2''=c_2'=[cos(\frac{\displaystyle{\theta_4}}{\displaystyle{2}})]^{1/N},
&s_2''=s_2'=[sin(\frac{\displaystyle{\theta_4}}{\displaystyle{2}})]^{1/N}\\[2mm]
c_2'''=c_3'=[cos(\frac{\displaystyle{\theta_5}}{\displaystyle{2}})]^{1/N},
&s_2'''=s_3'=[sin(\frac{\displaystyle{\theta_5}}{\displaystyle{2}})]^{1/N}\\[2mm]
c_3'''=c_3''=[cos(\frac{\displaystyle{\theta_6}}{\displaystyle{2}})]^{1/N},
&s_3'''=s_3''=[sin(\frac{\displaystyle{\theta_6}}{\displaystyle{2}})]^{1/N}\\
\end{array}\end{equation}
In this way, we have
\begin{equation}
u_i=\omega^{-1/2}[ctg(\frac{\theta_i}{2})]^{2/N}, ~~~~~~i=1,2,\cdots,6
\end{equation}
The vertex type tetrahedron equation has the form (see Figs. 3,4):
$$
{\displaystyle\sum_{\{k_i\},\atop i=1,\cdots,6}}
R(\theta_1,\theta_2,\theta_3)^{k_1,k_2,k_3}_{i_1,i_2,i_3}
R(\theta_1,\theta_4,\theta_5)^{j_1k_4k_5}_{k_1i_4i_5}
R(\theta_2,\theta_4,\theta_6)^{j_2j_4k_6}_{k_2k_4i_6}
R(\theta_3,\theta_5,\theta_6)^{j_3j_5j_6}_{k_3k_5k_6}= ~~~~~~~
$$
\begin{equation}
{}~{\displaystyle\sum_{\{k_i\},\atop i=1,\cdots,6}}
R(\theta_3,\theta_5,\theta_6)^{k_3,k_5,k_6}_{i_3,i_5,i_6}
R(\theta_2,\theta_4,\theta_6)^{k_2k_4j_6}_{i_2i_4k_6}
R(\theta_1,\theta_4,\theta_5)^{k_1j_4j_5}_{i_1k_4k_5}
R(\theta_1,\theta_2,\theta_3)^{j_1j_2j_3}_{k_1k_2k_3}
\end{equation}
with the angles satisfying the condition
$$
\bigg[sin\frac{\theta_1+\theta_2+\theta_3}{2}
sin\frac{-\theta_1+\theta_2+\theta_3}{2}
sin\frac{-\theta_3+\theta_5+\theta_6}{2}
sin\frac{\theta_3+\theta_5-\theta_6}{2}\bigg]^{1/2}~~~~~~~~~~~~
$$$$
-\bigg[sin\frac{\theta_1-\theta_2+\theta_3}{2}
sin\frac{\theta_1+\theta_2-\theta_3}{2}
sin\frac{\theta_3-\theta_5+\theta_6}{2}
sin\frac{\theta_3+\theta_5+\theta_6}{2}\bigg]^{1/2}~~~~~
$$
\begin{equation}\label{3.2}
{}~~~~~~~~~~~~~~~~~~~~~~~~~~~~~~~~~~~~~~=sin\theta_3\bigg[sin\frac
{\theta_2+\theta_4-\theta_6}{2}
sin\frac{-\theta_2+\theta_4+\theta_6}{2}\bigg]^{1/2}
\end{equation}
(see Fig. 5). This relation can be obtained from Eq. (3.2) of Ref.
\cite{Zam1} by a proper choice of the angles: $\theta_2\rightarrow
\theta_3,~\theta_3\rightarrow\pi-\theta_2,~\theta_5\rightarrow\theta_6,
{}~\theta_6\rightarrow\pi-\theta_5$. And the vertex type weight function has
 the property
\begin{equation}
R(\theta_1,\theta_2,\theta_3)^{j_1j_2j_3}_{i_1i_2i_3}=R(\theta_1,\theta_2,
\theta_3)^{i_1i_2i_3}_{j_1j_2j_3}
\end{equation}
The properties of the others of the weight functions will be given in the
following section. These angles $\theta_1, \theta_2, \cdots, \theta_6$ can
be
interpreted as the parameters related to the six spaces in which the vertex
tetrahedron equation is defined.

\section{\bf The Symmetry Properties of the Vertex Type
 Weight Function }

In this section we first consider the additional constraints imposed on the
 tetrahedron equations, given by Kashaev $et~al$, from the point of the
 above angle variables. Then we find that the Boltzmann weights are
symmetrical under the transformations of the group $G$ consisting of
 various rotations, reflections and their combinations of the cube in
the respect of the vertex type weight functions. It can be checked easily
 that the angle parametrization satisfies the conditions which ensures
all the similarity transformation factors \cite{Kaev1} to cancel each
 other. In terms of the `coordinated' parameters the four additional
constraints have the form
\begin{equation}\begin{array}{rr}
\omega\frac{\displaystyle{x_{23}}}{\displaystyle{x_3}}
\frac{\displaystyle{x_4'}}{\displaystyle{x_{24}'}}
\frac{\displaystyle{x_{24}''}}{\displaystyle{x_2''}}
\frac{\displaystyle{x_2'''}}{\displaystyle{x_{24}'''}}=1,&
\frac{\displaystyle{x_{13}}}{\displaystyle{x_1}}
\frac{\displaystyle{x_1'}}{\displaystyle{x_{14}'}}
\frac{\displaystyle{x_{14}''}}{\displaystyle{x_1''}}
\frac{\displaystyle{x_1'''}}{\displaystyle{x_{14}'''}}=1\\[6mm]
\frac{\displaystyle{x_{14}}}{\displaystyle{x_4}}
\frac{\displaystyle{x_4'}}{\displaystyle{x_{14}'}}
\frac{\displaystyle{x_{14}''}}{\displaystyle{x_4''}}
\frac{\displaystyle{x_4'''}}{\displaystyle{x_{24}'''}}=1,&
\frac{\displaystyle{x_{13}}}{\displaystyle{x^{\*}_3}}
\frac{\displaystyle{x_3'}}{\displaystyle{x_{13}'}}
\frac{\displaystyle{x_{13}''}}{\displaystyle{x_1''}}
\frac{\displaystyle{x_2'''}}{\displaystyle{x_{23}'''}}=1
\end{array}\end{equation}
By taking account of the expressions (28-30), the above constraints
 are changed into the relations
\begin{equation}\begin{array}{ll}
l_{12}+l_{24}=l_{14},&l_{13}+l_{35}=l_{15}\\[4mm]
l_{23}+l_{36}=l_{26},&l_{45}+l_{56}=l_{46}
\end{array}\end{equation}
Just as showing in Fig. 5, they hold naturally. From Refs.
 \cite{BB2,hu1}, we know that the three-dimensional star-star relation
 means the transformation $\xi$:
\begin{equation}
W(a|efg|bcd|h)\stackrel{\xi}{\longrightarrow} W(f|adb|hge|c)
\end{equation}
with
\begin{equation}
\frac{x_3}{x_1}\stackrel{\xi}{\rightarrow}\frac{x_2}{x_3},~~~
\frac{x_4}{x_2}\stackrel{\xi}{\rightarrow}\frac{x_1}{\omega x_4},~~~
\frac{x_4}{x_1}\stackrel{\xi}{\rightarrow}\frac{x_1}{\omega x_3},~~~
\frac{x_3}{\omega x_2}\stackrel{\xi}{\rightarrow}\frac{x_2}{\omega x_4}
\end{equation}
In terms of the vertex form the star-star relation can be expressed as
\begin{equation}\label{sy1}
R(\theta_1,\theta_2,\theta_3)^{j_1j_2j_3}_{i_1i_2i_3}=R(\pi-\theta_3,
\pi-\theta_2,\theta_1)^{-j_3-j_2i_1}_{-i_3-i_2j_1}
\end{equation}
as in Fig. 6. Under the transformations $\tau$ and $\rho$ of the generating
elements of the group $G$ the weight functions change as \cite{Newseies}
\begin{equation}
W(a|efg|bcd|h)\stackrel{\tau}{\longrightarrow} W(a|feg|cbd|h)
\end{equation}
with
\begin{equation}
\frac{x_2}{x_1}\stackrel{\tau}{\longleftrightarrow}\frac{x_3}{\omega x_4},~
{}~~~
\frac{x_4}{x_1}\stackrel{\tau}{\longleftrightarrow}\frac{x_3}{\omega x_2}
\end{equation}
and
\begin{equation}
W(a|efg|bcd|h)\stackrel{\rho}{\longrightarrow} W(g|cab|fhe|d)
\end{equation}
with
\begin{equation}
\frac{x_3}{x_1}\stackrel{\rho}{\rightarrow}\frac{x_{13}x_4}{x_3x_{14}},~~
\frac{x_3}{x_2}\stackrel{\rho}{\rightarrow}\frac{\omega x_{23}x_4}
{x_3x_{24}},~~
\frac{x_4}{x_2}\stackrel{\rho}{\rightarrow}\frac{x_{23}}{x_{24}},~~
\frac{x_{14}x_{23}}{x_{13}x_{24}}\stackrel{\rho}{\rightarrow}\frac{x_2}{x_1}
\end{equation}
So the vertex forms of them are
\begin{equation} \label{sy2}
R(\theta_1,\theta_2,\theta_3)^{j_1j_2j_3}_{i_1i_2i_3}=
R(\theta_3,\theta_2,\theta_1)^{j_3j_2j_1}_{i_3i_2i_1}
\end{equation}
for transformation $\tau$, as in Fig. 7, and
\begin{equation}\label{sy3}
R(\theta_1,\theta_2,\theta_3)^{j_1j_2j_3}_{i_1i_2i_3}=
R(\pi-\theta_2,\theta_1,\pi-\theta_3)^{-i_2j_1-j_3}_{-j_2i_1-i_3}
\end{equation}
for transformation $\rho$ as in Fig. 8. As is Known, the angles $\theta_1,
 \theta_2, \cdots, \theta_6$ can be interpreted as the dihedral angles
between the rapidity planes  connected  with the cubes.  In respect of the
vertex model, these angles  can be regarded as the parameters related to
the spaces on which the vertex type tetrahedron equation is defined. So
 these parameters and spin variables should be transformed `regularly'
under the symmetry group $G$. This is entirely consistent with the above
equations. The geometric considerations of them are shown in Figs 6, 7, 8.
The above two relations are the ``elementary'' relations. The others of the
 transformations of $G$ can be obtained  from them. It can be checked
 easily that the star-star relation (\ref{sy1}) can be obtained from
relations (\ref{sy2}) and (\ref{sy3}).

\section{\bf Summary}

As the conclusions, we get the duality between the cube type weight
functions and the vertex type weight functions explicitly for the
three-dimensional Baxter-Bazhanov model and find that the Boltzmann
 weight of the model depends on the four spin variables which are the
linear combinations of the spins located on the corner sites of the
 cube. We can interpreted the vertex type weight function $R(\theta_1,
\theta_2,\theta_3)^{j_1j_2j_3}_{i_1i_2i_3}$ as a Boltzmann weight of a
three-dimensional vertex model and the spectrums $\theta_1,\theta_2,
\theta_3$ are connected to the ``line'' $1,2$ and $3$ (see Fig. 2).
 In this way, we write down the symmetrical relations of the vertex type
 Boltzmann weights in terms of the angles. We known that these angles
 are the dihedral angles between the rapidity planes connected with the
cubes. Then the weight functions should be transformed `regularly' under
the actions of the symmetry group $G$ which is consisted by the various
 rotations, reflections and their combinations of the cube. The relations
(39), (44) and (45) are  entirely consistent with it (see Fig. 6,7,8).
And the angles $\theta_1,\theta_2,\theta_3,\theta_4,\theta_5,\theta_6$
 with the relation (\ref{3.2}) can be interpreted as the spectrums related
 to the six spaces in which the vertex tetrahedron equation (32) is
defined (see Fig. 3 and Fig. 4). When we set $i_3\equiv j_3\equiv 0$
and make the specialization of the spectral parameters the  Boltzmann
 weight of the vertex model proposed in Ref.\cite{newsolu} can be obtained
from the  weight function (19) with the spin assignments \cite{hunew}.
 Here we have 16 nonzero weights
$R(\theta_1,\theta_2,\theta_3)^{j_1j_2j_3}_{i_1i_2i_3}$ for $N=2$.
 So it is a interesting question to find the connection between it and
the solutions in Ref. \cite{Kor}.

\section*{\bf Acknowledgment}

One of the authors (Hu) would like to thank K. Wu for the interesting
discussions and J. X. Liu, Z. B. Su for the encouragements during this
 work. I (Hu) gratefully acknowledge Y. K. Zhou for taking me to the
 field of statistical mechanics.

\newpage
\section*{\bf Captions}

Fig. 1. Arrangements of the spins $a,\cdots,h$ on the corner sites and
the spin $\sigma$ in the
 \* \* \* \* \* \* \* \* \* \*  \*  center of an elementary cube of the
simple cubic lattice ${\cal L}$.

\noindent
Fig. 2. The Boltzmann weight of the three-dimensional vertex model
 corresponding
 \* \* \* \* \* \* \* \* \* \* \*   to the IRC model.

\noindent
Fig. 3. The graph of the left hand sides of the tetrahedron equations.

\noindent
Fig. 4. The graph of the right hand sides of the tetrahedron equations.

\noindent
Fig. 5. A fragment of the stereo-graphic projection of the sphere with
four great
 \* \* \* \* \* \* \* \* \* \*  \*  \* \*  \*    circles.

\noindent
Fig. 6. The transformation $\xi$ corresponding to the three-dimensional
 star-star
 \* \* \* \* \* \* \* \* \* \*  \*  \* \*  \* \* \*  \*  \*  relation.

\noindent
Fig. 7  The transformation $\tau$ of the weight function.

\noindent
Fig. 8  The transformation $\rho$ of the weight function.

\end{document}